\documentclass[aps,prl,twocolumn,showpacs,superscriptaddress,groupedaddress]{revtex4}
\usepackage{amsmath}
\usepackage{graphicx}
\usepackage{natbib}
\usepackage[usenames]{color}
\usepackage[T1]{fontenc}
\usepackage{textcomp}

\begin{document}


\title{Analytical models for well-mixed populations of cooperators and defectors under limiting resources}

\author{R.~J.~Requejo}
\author{J.~Camacho}
\affiliation{Departament de F\'isica, Universitat Aut\`onoma de Barcelona, Campus UAB, E-08193 Bellaterra, Spain.}
\email {juan.camacho@uab.es}

\date{\today}

\begin{abstract}
In the study of the evolution of cooperation, resource limitations are usually assumed just to provide a finite population size. Recently, however, agent-based models have pointed out that resource limitation may modify the original structure of the interactions and allow for the survival of unconditional cooperators in well-mixed populations. Here we present analytical simplified versions of two types of agent-based models recently published: one in which the limiting resource constraints the ability of reproduction of individuals but not its survival, and a second one where the limiting resource is necessary for both reproduction and survival. One finds that the analytical models display, with a few differences, the same qualitative behavior of the more complex agent-based models. In addition, the analytical models allow to expand the study and identify the dimensionless parameters governing the final fate of the system, such as coexistence of cooperators and defectors, or dominance of defectors or of cooperators. We provide a detailed analysis of the occurring phase transitions as these parameters are varied.  
\end{abstract}


\maketitle

Cooperation is common in nature and it is considered to have played a key role in the evolutionary appearance of higher selective units, such as eukaryotic cells or multicellular life, from simpler components \cite{maynard-smith:1995a}. However, its abundance is intriguing because cooperators are vulnerable to exploitation by defectors \cite{darwin:1859,hamilton:1964a,hamilton:1964b}. Several mechanisms have been found allowing cooperative behaviors to survive such as network structure, group selection, direct and indirect reciprocity or tag-based donation \cite{riolo:2001,nowak:2006b}. Behavioral mechanisms --such as the latter three examples-- require players to have some ability to avoid the exploitation from defectors, such as memory or the capacity to recognise the co-player \cite{riolo:2001,nowak:2006b}. As a consequence, simple agents without these abilities, such as unconditional cooperators, are not expected to survive in well-mixed populations. 


Aside from a few examples \cite{wakano:2007,dobramysl:2008,hauert:2006a,melbinger:2010}, the role played by the limitation of resources in most studies on the origin and persistence of cooperation has been just to impose a constant population size \cite{trivers:1971,axelrod:1981,nowak:1998,nowak:2006b,riolo:2001,traulsen:2003,gomez-gardenes:2007,roca:2009}. Recently, however, we have introduced a new viewpoint where the environment is considered explicitly by introducing a resource flux into the system that drives it away from equilibrium \cite{requejo:2011,requejo:2012}. This standpoint leads to unexpected outcomes, such as that resource limitation allows for stable coexistence between unconditional cooperators and defectors, and even dominance of cooperation, in well-mixed populations playing an a priori Prisoner's Dilemma (PD) game, where defectors are expected to drive cooperators to extinction \cite{axelrod:1981}. This happens due to a selforganizing process involving the environment which generates dynamical payoffs transforming the original PD structure into a different game. 


Two main scenarios can be devised for the limiting resource: (a) the latter is necessary for reproduction but not for survival (such as for instance, any substance required for the development of the embryo) as analyzed in \cite{requejo:2012} or (b) it is necessary for both reproduction and survival (such as sugars and some other nutrients), as analyzed in \cite{requejo:2011}. In both cases, agent-based models showed that coooperation could survive for some parameter values. 
The purpose of this paper is to develope simplified analytical models capturing the mechanisms involved in the agent-based models. This will supply a more complete view of the model behavior and will provide the dimensionless groups ruling the dynamics. We show that the analytical models yield a similar qualitative behavior of the agent-based models, though with a few differences. The paper is organized as follows. In section I, we analize the case where the limiting resource constrains reproduction but not survival. We present the model, its corresponding equations and analize with detail their solutions and phase transitions. Section II performes a similar analysis for the case where the limiting resource constrains both reproduction and survival. Section III is devoted to conclusions.

\section{Limiting resource constraining reproduction}

\subsection{Simplified analytical model}

We develop here an analytical model aimimg to mimic the model of ref. \cite{requejo:2012}. For the paper to be selfcontained, we provide a summary of the latter model. It consists of an evolving well-mixed population of self-replicating individuals that receive resources from the environment and exchange resources during interactions. No memory, learning abilities or any other sensory inputs are assumed. Each individual $i$ is represented by its internal amount of resources $E_i$ and its strategy, namely cooperate (C) or defect (D). The internal amount of resources may be interpreted as the amount that belongs to it, independently of why or how. The environment provides resources in portions $E_{0}$ per unit time to randomly chosen individuals independently of their strategy, thus not modifying the structure of the payoffs; the total resource influx $E_T$ is assumed to be constant, so that $E_{0}=E_T/N$, with $N$ the population number. If the amount of resources of an individual exceeds a value $E_s$, it splits into two identical copies with half its internal amount of resources. Deaths occur with probability $f$ per unit time. Defectors are characterized by the maximum amount of resources associated to an interaction: the cost spent $(E_c)$ for stealing a reward $(E_r)$ from the co-player. If the internal resources of a defector are smaller than the cost $E_c$, it does not pay the cost nor receives the reward. If the interaction partner has resources below the reward, the entire amount of resources is sequestered. Thus, the quantities associated to the defector strategy represent the ideal outcome of the interactions. We assume that these quantities are inherited without mutation. Large populations with $E_r>E_c>0$ were considered, so that the system plays a priori a PD game \cite{requejo:2012}.

In the simplified model, the internal amount of resources $E_i$ is either 0 or 1. Each defector attacks at a rate $\alpha$ per unit time to individuals chosen at random 
and steals its internal resources. To do so, the defector must have internal resources greater than 0 (i.e. $E_i=1$), otherwise it does not attack. In every interaction, the defector loses its unit of resources with probability $q$, which can thus be seen as the average cost paid by a defector in an interaction. 
If the interaction partner has no resources, no reward is obtained. Cooperators do nothing, they just eventually suffer from defector's attacks. Let us note that, in an ideal situation where all individuals possess resources ($E_i=1$), 
the payoffs for cooperators and defectors in an interaction between them are respectively $ f_{CD}=-1, f_{DC}=1-q$; interactions between cooperators result in a payoff $f_{CC}=0$ for each and between defectors in $f_{DD}=-q$. The payoff ordering in this situation $ f_{DC}>f_{CC}>f_{DD}>f_{CD}$ corresponds to a Prisoner's Dilemma, and hence the names cooperators and defectors, and the expectation of the extinction of cooperation. However, since not all cooperators have resources $E_i=1$, the average net reward got by defectors will be, in general, smaller than $1-q$. This reward depends on the fraction of cooperators in state $E_i=1$, which is a dynamical quantity. As a consequence, the dynamics might lead the average reward to values smaller or equal to the cost and allow for the survival of cooperators, as we will see below.

The system receives from the environment $E_T$ units of resources per unit time, which are distributed equally among the $N$ individuals of the population independently of its strategy, thus not modifying the interaction payoff structure. When an individual with internal resources $E_i=1$ receives an extra unit of resources it splits into two identical copies, each one with $E_i=1$. Along with reproduction, we assume that players die with a probability $f$ per unit time, independently of its strategy. Therefore, resource allocation, reproduction and death rules are equal for both cooperators and defectors, being the strategy the only difference.

At the sight of the two models, the main differences between them are that in the simplified model the resource distribution is discrete (while it is continuous in the agent-based model), and defectors cost is stochastic.

We consider simultaneous interactions and large populations so that we can make a continuum approach. We denote by $c_0$ and $c_1$ the number of cooperators with internal resources 0 and 1, $d_1$ and $d_0$ the number of defectors with internal resources 1 and 0, respectively; $N=c_0+c_1+d_1+d_0$ is then the total population size. The evolution equations according to the mechanisms involved are
\begin{eqnarray}
\label{c1}
\frac{dc_0}{dt}&=& \alpha d_1 \frac{c_1}{N} - \frac{E_T}{N} c_0 - f c_0 \\
\label{c2}
\frac{dc_1}{dt}&=& - \alpha d_1 \frac{c_1}{N}+ \frac{E_T}{N} (c_0 + c_1) - f c_1 \\ \nonumber
\label{d1}
\frac{d d_1}{dt}&=& \alpha (1-q) d_1 \frac{c_1}{N}- \alpha q d_1(\frac{c_0+d_0+d_1}{N})+ \\ 
&+&\frac{E_T}{N} (d_0 + d_1) - f d_1 \\ \nonumber
\label{d0}
\frac{d d_0}{dt}&=& \alpha (1-q) d_1 \frac{d_1}{N}+ \alpha q d_1(\frac{c_0+d_0+d_1}{N})- \\ 
&-& \frac{E_T}{N} d_0 - f d_0 
\end{eqnarray}
Let us first explain the interaction terms. From the $\alpha$ attacks per unit time of a defector $D_1$, in a fraction $c_1/N$the victim will be a $C_1$ player. Thus, $\alpha d_1 c_1/N$ describes the rate of interactions of active defectors $D_1$ with cooperators $C_1$. As a result of these interactions, individuals $C_1$ lose their internal resource unit after the attack of defectors and move from population $c_1$ to $c_0$; this explains the first term in the equations for $c_0$ and $c_1$. Also, in a fraction $1-q$ of these interactions the $D_1$ individual keeps its resource unit, which added to the stolen unit from the $C_1$ player, leads to its reproduction; this gives the first term in the equation for $d_1$. When a $D_1$ player attacks $C_0$ and $D_0$ individuals, a fraction $q$ of times loses its resource unit so that it moves to population $D_0$ (this yields part of the second term in equations for $d_1$ and $d_0$). Finally, an interaction $D_1 D_1$ produces either $D_1 D_0$ with probability $q$ (which reduces the $D_1$ population and increases $D_0$), or $D_1 D_1 D_0$ with probability $1-q$, which does not affect the population of $d_1$ but increases the one of $d_0$; these two processes describe the remaining interaction terms in the dynamic equations for $d_1$ and $d_0$.

On the other hand, the term in $\frac{E_T}{N} c_0$ quantifies the number of individuals $C_0$ moving to population $C_1$ after getting a unit of resources from the environment. In addition, individuals in population $c_1$ that receive resources from the environment replicate, thus increasing the $c_1$ population. The same applies for the population of defectors $d_0$ and $d_1$. The terms with $f$ just describe the number of individuals dying in each population per unit time. 

This model is much simpler than the one described in \cite{requejo:2012}, because it contains just four independent variables, namely $c_0, c_1, d_1$ and $d_0$, in contrast to the many variables included in the resource distributions for cooperators and defectors of the latter model. However, it keeps most of its ingredients and captures the main features of its behaviour: the promotion of cooperation triggered by resource limitations.

\subsection{Model dynamics}

\begin{figure*}
	\centering
	 \includegraphics[width=88mm]{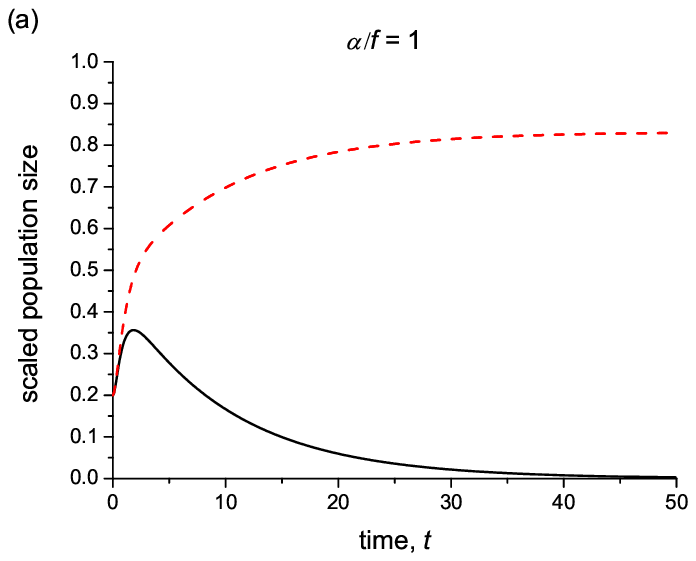} \includegraphics[width=88mm]{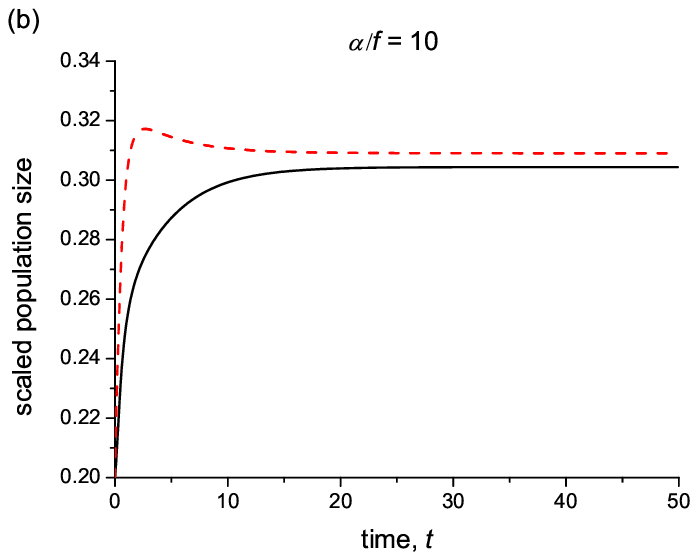}
\caption{\label{fig1} 
Time evolution for the scaled number of cooperators (solid line) and number of defectors (dashed line) for two sets of parameter values: $q=0.7$ and (a) $\alpha/f=1$ and (b) $\alpha/f=10$. In (a) cooperators die out, in (b) the system ends up in a mixed state where cooperators and defectors coexist. These two types of states are the only attractors of the dynamics. The populations are given in units of $E_T/f$ (see Eq. (\ref{dim})).}
\end{figure*}

One of the properties of the model in \cite{requejo:2012} is that a change in the resource influx $E_T$ does not modify the final fate of the system, but just the final population size; more specifically, it was found in equilibrium that $N \propto E_T$. This is also what the system (\ref{c1})--(\ref{d0}) predicts for the steady state, since the right-hand side of all the equations are homogeneous functions of degree 1 of variables $c_i, d_i, N$ and $E_T$. This means that the stationary states are solutions of the type $c_i, d_i= \lambda_i E_T$, with $\lambda_i$ constants depending only on $\alpha, f$ and $q$; therefore, $N \propto E_T$ as in the previous model. 

One can obtain a further understanding of the model through nondimensionalization. By dividing Eqs.(\ref{c1})--(\ref{d0}) by the death rate $f$ and defining the dimensionless time $ft$, the system turns to be described by three dimensionless parameters, namely $q$, $\alpha/f$ and $E_T/f$. Since we have seen above that stationary populations are proportional to $E_T$, they can be generally written as
\begin{equation}
\label{dim}
c_i,d_i=g_i(q,\frac{\alpha}{f})\frac{E_T}{f},
\end{equation}
with $g_i$ unknown functions of the dimensionless parameters $q$ and $\alpha/f$. Then, $E_T/f$ sets the characteristic size of the populations, and the composition, let us say the population fractions $c_i/N, d_i/N$, comes determined by the other two parameters, $q$ and $\alpha/f$ through functions $g_i$. Parameter $\alpha/f$ has a direct interpretation: since $\alpha$ is the attack rate of a defector and $f^{-1}$ is the average lifetime of an individual, $\alpha/f$ denotes the average number of interactions performed by an active defector, and it is this quantity, along with the cost $q$, what determines the fate of the system. For simplicity, in the following we will assume $E_T=1$, i.e. absolute populations $c_i,d_i$ will be given in units of $E_T$.

The most interesting feature of the model in \cite{requejo:2012} was the existence of situations where cooperators are able to survive in coexistence with defectors in mixed stationary states; otherwise the population reaches a stationary state with only defectors. We analyze next the dynamical behavior of Eqs. (\ref{c1})--(\ref{d0}) through its numerical and analytical resolution.  
Figure \ref{fig1} illustrates that the dynamic equations (\ref{c1})--(\ref{d0}) indeed display a similar behavior as the original model, namely by modifying parameters $\alpha, f$ and $q$ the system ends up either in purely defective states or in coexistence states.

\begin{figure*}
	\centering
	 \includegraphics[width=88mm]{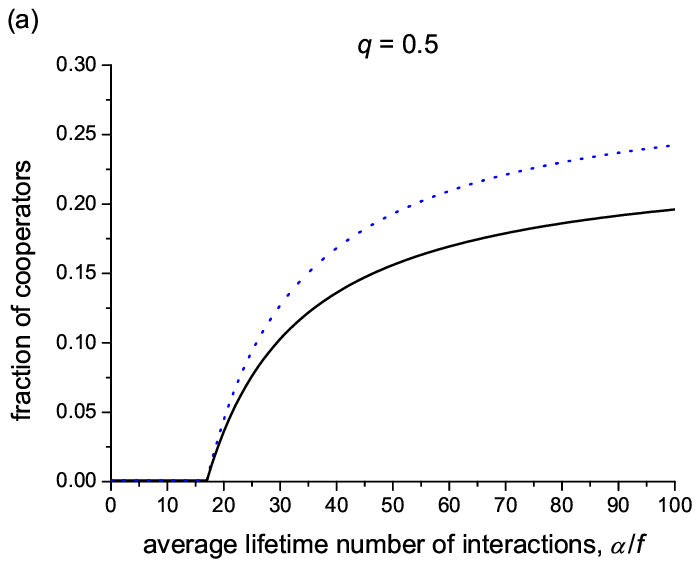} \includegraphics[width=88mm]{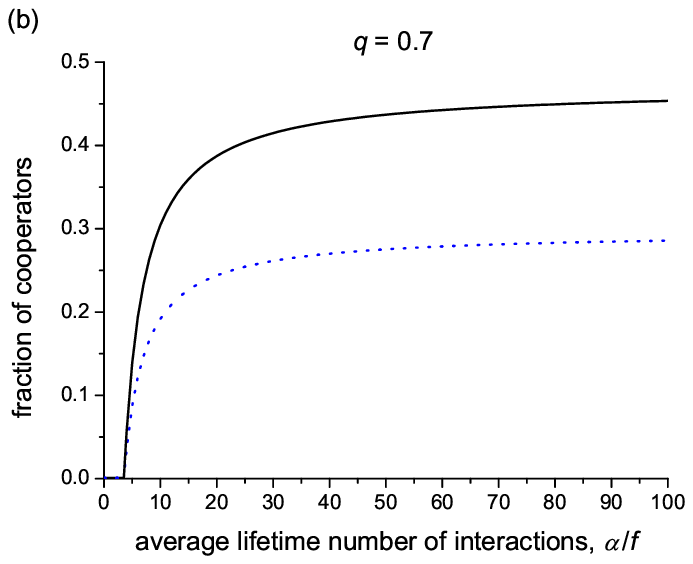}
\caption{\label{fig2} 
Phase transition. Fraction of cooperators ($c_1/N$, solid line; $c_0/N$, dotted line) as a function of parameter $\alpha/f$ for (a) $q=0.5$ and (b) $q=0.7$. Below a critical value, cooperators die out.}
\end{figure*}
\begin{figure*}
	\centering
	 \includegraphics[width=88mm]{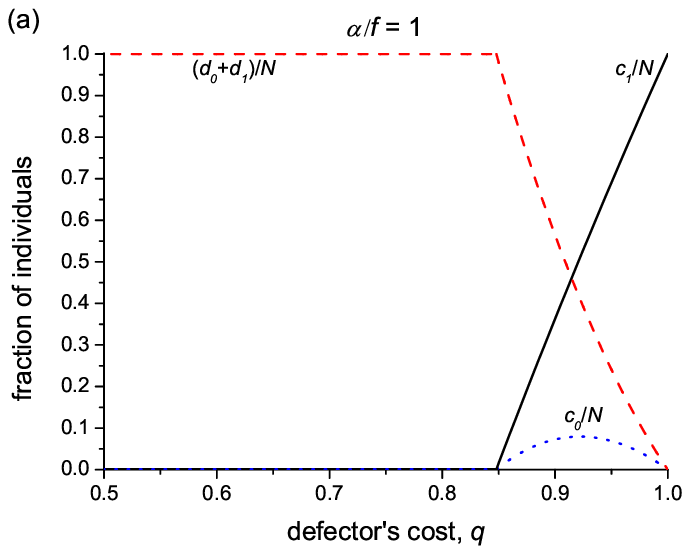} \includegraphics[width=88mm]{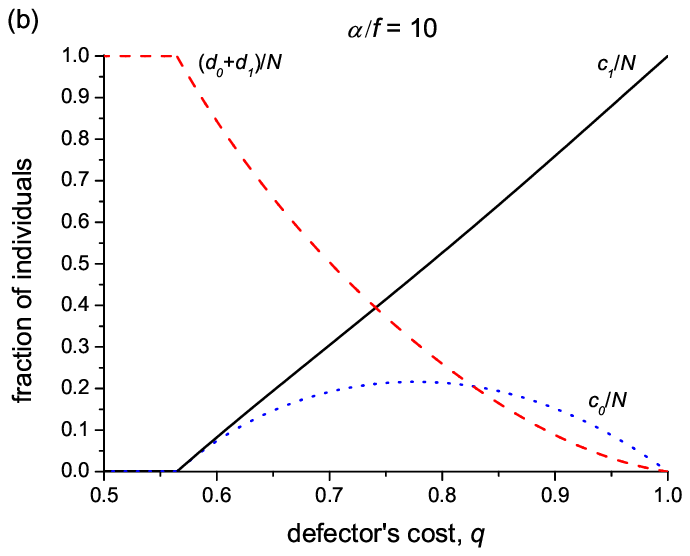}
\caption{\label{fig3} 
Phase transition. Fraction of cooperators ($c_1/N$, solid line; $c_0/N$, dotted line) and defectors (dashed line) as a function of parameter $q$ for (a) $\alpha/f=1$ and (b) $\alpha/f=10$. Below a critical value, cooperators die out.}
\end{figure*}

One can perform a systematic analysis of the dynamic behavior of our system (\ref{c1})--(\ref{d0}) which will allow for a better understanding of the model. Its analytical resolution shows the existence of a number of stationary states: (a) one made only of cooperators $C_1$, which the dynamics shows it is always unstable; (b) another one made only of defectors, which is sometimes stable (we give it here for completeness, $d_1=\frac{1}{\alpha q+f}$, $d_0=\frac{\sqrt{1+\alpha/f}-1}{\alpha q+f}$); and (c) a number of mixed states of which only one provides positive values for all population variables for some parameters. The latter is the mixed state that, when takes positive values, becomes stable at the same time that the defective state (b) becomes unstable. In order to study this (transcritical) transition, it is useful to realise that the populations for cooperators in the mixed state (c) have the following form (the solutions for defectors are always positive) 
\begin{equation}
c_i=\frac{a_i}{\alpha}[\frac{\alpha}{f}-\gamma_c],
\label{ci}
\end{equation}
where $a_i$ $(i=0,1)$ and $\gamma_c$ are functions of $q$ only. This expression supplies positive values for $c_i$ only when $\alpha/f > \gamma_c$. Therefore, $\gamma_c(q)$ is the critical value over which $\alpha/f$ must be in order for the system to end up into a coexistence state. This indicates that the transition from defective to coexistence states depends on the parameter $\alpha/f$, and not separately of parameters $\alpha$ and $f$, in agreement with Eq. (\ref{dim}). 
Fig.\ref{fig2} shows the phase transition from dominance of defectors to coexistence of cooperators and defectors for two values of $q$.
One observes that the survival of cooperation is favored by larger defector costs (larger $q$), as expected, since mixed states appear at lower critical numbers of $\alpha/f$. 
One can also display the phase transition as a function of $q$ for two values of $\alpha/f$ (Fig. \ref{fig3}). One observes that as $q \rightarrow 1$ the mixed state tends to the stationary state made only of cooperators $C_1$, which is the stable state for $q=1$. 
Finally, one can obtain the phase diagram separating the two behaviors in terms of the two parameters governing the system. The separation line is found by imposing $c_i=0$ in Eq. (\ref{ci}), i.e.  $\alpha/f=\gamma_c$. The analytical expression for  $\gamma_c$ is long but it can be solved numerically (see Fig. \ref{fig4}). 

\subsection{Comparison with the agent-based model}
\begin{figure}
	\centering
   \includegraphics[width=88mm]{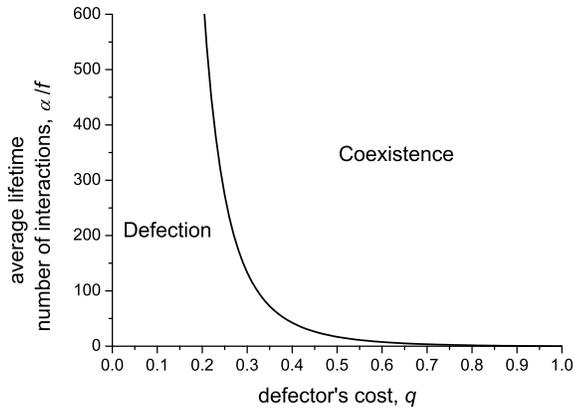}
\caption{\label{fig4} 
Phase diagram. For large costs $q$ or large average number of attacks in a lifetime, $\alpha/f$, cooperators survive in coexistence with defectors.}
\end{figure}

The phase diagram, Fig. \ref{fig4}, shows that cooperation is favored at large defectors' costs $q$, as expected, and also at large numbers of attacks in a lifetime, $\alpha/f$. The latter behavior is surprising, since at first sight interactions should benefit defectors versus cooperators. However, we must bear in mind that attacks are indiscriminate and then a fraction of the attacks fall on defectors themselves, thus reducing the number of defectors in active states. The dynamics of the system shows that for a large enough number of attacks in a lifetime, the number of active defectors decrease enough so as to allow for the survival of cooperators. This happens in a continuous phase transition as seen in Figs. \ref{fig2} and \ref{fig3}. Let us note that the latter behaviour was partially observed in the original agent-based model. There, the attacking rate $\alpha$ was set to unity by construction of the model, so that it could not be modified. However, it was found that cooperation increased when the death rate $f$ decreased, in agreement with the predictions of the present model. 

Let us further compare the agent-based model in \cite{requejo:2012} and the simplified analytical model presented here. As previously noticed, they show a similar qualitative behavior, both displaying a phase transition from purely defective states at low costs and small death rates to stable coexistence states in the opposite limit. However, there exist a few differences between them. The numerical simulations of the former model predicted dominance of cooperators for large costs (see Fig. \ref{fig2} in ref. \cite{requejo:2012}), whereas this does not occur in the analytical model. We have checked this difference of behavior by performing large-size simulations of the agent-based model, thus minimizing finite size effects; they confirm the extinction of defectors for large costs, and the existence of a region where they extinguish due to stochastic fluctuations in coexistence states with few defectors.



\section{Limiting resource constraining reproduction and survival}
In this section we provide an analytical model for the study of a well-mixed population of unconditional cooperators and defectors which exchange resources among them and with the environment but, in contrast to the previous section, here the resource limits not only the reproduction of individuals but also their survival. A more complex, agent-based model, analyzing this situation was presented in \cite{requejo:2011}. Again, the model consists of an evolving well-mixed population of self-replicating individuals that receive resources from the environment and exchange resources during interactions. Everything works as in the previous model except in two issues: (i) now individuals do not die at random, but when individuals exhaust their resources, and (ii) each individual dissipates an amount of resources $E_l$ per unit time in order to keep alive.

\subsection{Analytical model}

The simplified model follows the same spirit of the previous section. The internal resources of individuals are either 0 or 1. Each defector attacks at a rate $\alpha$ individuals chosen at random and steals its internal resources. To do so, the defector must have internal resources greater than 0 (i.e. $E_i=1$). In every interaction, the defector loses its unit of resources with probability $q$, which is thus the average cost paid by a defector in an interaction. The system receives resources from the environment at a rate $E_T$, and they are distributed equally among the $N$ individuals of the population independently of its strategy. In addition, individuals dissipate resources in living activities at a rate $r$; this is implemented as the probability of spending one resource unit per unit time. If an individual with 0 internal resources is attacked or it is required to dissipate resources it dies. On the other side, when an individual with internal resources $E_i=1$ receives an extra unit of resources it splits into two identical copies, each one with $E_i=1$. Again, resource allocation, reproduction and death rules are equal for both cooperators and defectors, being the strategy the only difference.

Let us note that the main differences of the model presented here and the latter model are again: (a) now there are only two resource levels $E_i=0,1$, instead of a much larger distribution, (b) the cost paid by active defectors in an interaction, and resource dissipation are stochastic. 

We use the notation of the previous section for the number of cooperators $c_i$ and defectors $d_i$. The model equations are very similar to them: 
\begin{eqnarray}
\label{c1s}
\frac{dc_0}{dt}&=& - \alpha d_1 \frac{c_0}{N}+\alpha d_1 \frac{c_1}{N} - \frac{E_T}{N} c_0 - r c_0 +r c_1\\
\label{c2s}
\frac{dc_1}{dt}&=& - \alpha d_1 \frac{c_1}{N}+ \frac{E_T}{N} (c_0 + c_1) - r c_1 \\ \nonumber
\label{d1s}
\frac{d d_1}{dt}&=& \alpha (1-q) d_1 \frac{c_1}{N}- \alpha q d_1(\frac{c_0+d_0+d_1}{N})+ \\ 
&+&\frac{E_T}{N} (d_0 + d_1) - r d_1 \\ \nonumber
\label{d0s}
\frac{d d_0}{dt}&=& - \alpha d_1 \frac{d_0}{N}+\alpha (1-q) d_1 \frac{d_1}{N}+ \alpha q d_1(\frac{c_0+d_0+d_1}{N})- \\ 
&-& \frac{E_T}{N} d_0 - r d_0 + r d_1 
\end{eqnarray}
The interaction terms are as in Eqs. (\ref{c1})--(\ref{d0}), with the exception of the first term in the  equations for $c_0$ and $d_0$, describing deaths due to interactions. In addition, dissipation kills individuals $c_0$ and $d_0$ at a rate $r$ and moves individuals from $c_1$ to $c_0$, and from $d_1$ to $d_0$ also at a rate $r$.

\subsection{Model dynamics}
One can analyze the dynamical behavior of our model equations (\ref{c1s})--(\ref{d0s}) following a scheme analogous to the one performed in the previous section. Since the right-hand side of the system equations are homogeneous functions of $c_i, d_i$ and $E_T$, the stationary populations $c_i, d_i$ are proportional to $E_T$. Therefore, $E_T$ just determines the population size, but not the composition. This is in agreement with the behavior found in the agent-based model \cite{requejo:2011}. One can also nondimensionalize the system equations by 
defining a nondimensional time $rt$. This shows that the system is ruled by three nondimensional parameters: $q$, $\alpha/r$ and $E_T/r$ and, analogously to the previos section, the stationary populations obey the scaling relation
\begin{equation}
\label{dims}
c_i,d_i=\overline{g}_i(q,\frac{\alpha}{r})\frac{E_T}{r}.
\end{equation}
Again, $E_T/r$ provides the characteristic size of populations $c_i,d_i$; this is larger the bigger the resource influx rate $E_T$ and the smaller its dissipation rate $r$. On the other hand, the composition of the final population is determined by parameters $q$ and $\alpha/r$. 

\begin{figure*}
	\centering
   \includegraphics[width=170mm]{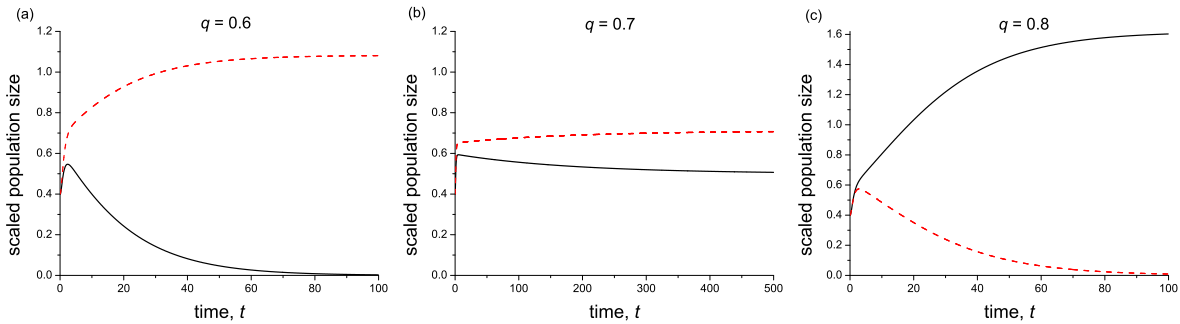}
\caption{\label{fig6} 
Time evolution for the scaled number of cooperators (solid line) and number of defectors (dashed line) for three sets of parameter values: $\alpha/r=1$ and (a) $q=0.6$, (b) $q=0.7$ and (c) $q=0.8$. In (a) cooperators die out, in (b) the system ends up in a mixed state where cooperators and defectors coexist, in (c) cooperators get rid of defectors. Two phase transitions occur as $q$ increases. The populations are given in units of $E_T/r$ (see Eq. (\ref{dims})).}
\end{figure*}

\begin{figure}
	\centering
   \includegraphics[width=88mm]{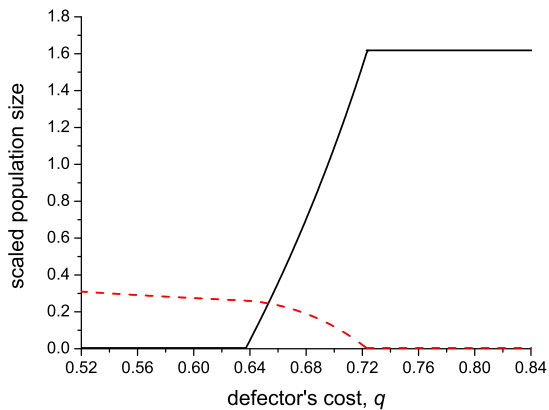}
\caption{\label{fig7} 
Phase transitions for $\alpha/r=10$. Population of defectors (dashed line) and cooperators (solid line) as function of defectors cost $q$. As $q$ grows two phase transitions occur: first, a transition from defective to coexistence states, and a second one at $q=0.723$ from coexistence to purely cooperator states. }
\end{figure}

The numerical resolution of the system (\ref{c1s})--(\ref{d0s}) displays two main behaviors depending on the parameter values: dominance of defection and dominance of cooperation (see Fig. \ref{fig6}), in agreement with the behavior of the agent-based model \cite{requejo:2011}, and in contrast to the case where the limiting resources just affect reproduction and not survival discussed in the previous section. Fig. \ref{fig6}b shows that coexistence of cooperators and defectors is also possible, in oposition to the behavior of the more complex agent-based model. Again, Fig. \ref{fig6} shows that the final composition of the system changes as a parameter is varied, indicating a phase transition from a population of only defectors at small costs $q$ and a population of only cooperators at larger $q$ values, separated by a coexistence region (see Fig. \ref{fig7}). Let's analyze the parameter region where each attractor becomes the stable one, i.e. let us find the phase diagram of the model.

\begin{figure}
	\centering
   \includegraphics[width=88mm]{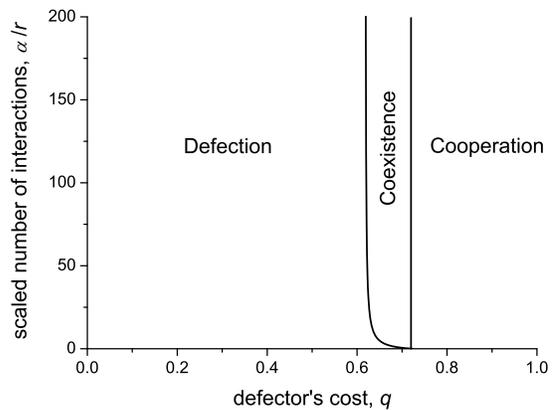}
\caption{\label{fig8} 
Phase diagram. Below $q=0.62$ defectors dominate. Above $q=0.72$ cooperators dominate. In between there is a thin region of coexistence of cooperators and defectors.}
\end{figure}

The analytical resolution of the system reveals the existence of three stationary states: (a) one made of cooperators ($c_1=E_T/r, c_0=0.62 E_T/r)$, (b) one made of defectors ($d_1=E_T/(\alpha q+r), d_0=E_T[\sqrt{\alpha^2+5r^2+6\alpha r}-r-\alpha]/2r(\alpha q+r))$, 
and (c) a mixed state whose solution provides positive solutions only for a limited set of parameter values. The whole solution is rather cumbersome, but one can extract valuable information by focussing in some aspects of it. The solution for $c_1$ is ($E_T=1$)
\begin{equation}
c_1=\frac{2.62\alpha q^2-1.62 q \alpha+3.62 q r-2.62 r}{q\alpha(1-q)},
\end{equation}
which is positive provided that
\begin{equation}
\label{phdg}
\frac{\alpha}{r}=\frac{1-1.38 q}{q(q-0.62)}>0.
\end{equation}
This shows that the mixed state can only exist in a thin range of $q$ values between 0.62 and 0.72, in agreement with Fig. \ref{fig6}b. The same happens for solution $c_0$. Populations $d_0$ and $d_1$ become negative at $q>0.72$. Expression (\ref{phdg}) provides the separation line between dominance of defectors and coexistence; for $q>0.72$ cooperators dominate (see Fig. \ref{fig7}). Therefore, the dynamics is mainly ruled by the defector cost $q$, being defectors dominant below $q=0.62$ and cooperators dominant above $q=0.72$ (Fig.\ref{fig8}). Aside from the thin range of coexistence in the parameter region separating both dominating behaviors, the analytical model thus provides the same qualitative behavior as the agent-based model presented in \cite{requejo:2011}.

\section{Conclusions} 
In this paper, we have studied the dynamics of two analytical models describing the evolution of well-mixed populations of unconditional cooperators and defectors under limiting resources. 
These models try to capture the essence of two separate agent-based models  dealing with two different situations. In one of them, the limiting resource restricts the ability of individuals to reproduce but do not alter their survival conditions. In the other, the resource rules both the survival and reproduction of individuals. The main differences of the models introduced here and the agent-based models are, on the one hand, that the distribution of internal resources is limited to two states, instead of a continuous distribution and, on the other, that the cost paid by defectors is now a stochastic process. 

Analytical models have the advantage with respect to simulation models of allowing for a complete and more compact analysis of their behaviour. Indeed, we have expanded the study of the agent-based models by separating the time scales of all the mechanisms in the model. In effect, in the agent-based models, we assumed equal rates for defectors' attacks and the feeding process. Here, in contrast, we consider different rates for each process. This increases in one the number of parameters. However, the nondimensionalization of the models has permitted us to identify the dimensionless parameters ruling the model dynamics, a study which is not easy to perform in agent-based models. As a result, we have easily seen that the resource influx from the environment determines the size of the final population, but not the composition of the population, a result that was observed in the simulations of the agent based models. The final composition is ruled by two parameters: the average cost paid by defectors and the number of attacks in a characteristic time.

The behaviour of the analytical models ressemble very much the ones obtained in the corresponding more complex agent-based models, remarkably allowing for the survival
of cooperators in some regions of the parameter space. 
When resources limit only reproduction, cooperators are able to coexist with defectors at larger defectors cost and larger number of interactions in a lifetime. The latter behavior may seem surprising, as one would expect defectors attacks indefectively to benefit defectors and harm cooperators. However, since attacks are indiscriminate, a large number of interactions in a lifetime reduces the number of defectors in active states and eventually allows for the survival of cooperators. Remarkably, this process occurs following a phase transition, so that cooperators are able to survive only when parameters surpass some critical value. When resources restrict reproduction and survival the fate of the system is essentially a population of defectors at small defector costs, and a population of cooperators at large costs, separated by a thin region of coexistence at intermediate costs. Aside from the coexistence region, this is the behavior found in the agent-based model. The origin of this difference is probably due to the different resource distribution in both models (discrete in the analytical model, and continuous in the agent based model). 


Finally, the work presented here could be extended by modifying the way resources are delivered and foraged by agents. If the mobility of agents is limited, these may have only the possibility to explore nearby resources and thus not all the resources provided by the environment would be used. Therefore it would be interesting to study the implications of the limitation of resources including space, and analysing inhomogeneous distributions of resources. This situation will probably enhance the survival of cooperators, but it may also lead to unexpected results. 




Fruitful discussions with X. Alvarez are acknowledged. This work has been supported by the Spanish government (FIS2009-13370-C02-01) and the Generalitat de Catalunya (2009SGR0164). R.J.R. acknowledges the financial support of the Universitat Autònoma de Barcelona and the Spanish government (FPU grant).

\end{document}